\documentstyle[12pt]{article}
\title{\bf Curvature expansion for the background--induced
gluodynamics string}
\author{D.V.ANTONOV \thanks{E-mail:
antonov@vxitep.itep.ru; supported by Deutsche Forschungsgemeinschaft
under contract 436 RUS 113/29, Russian Fundamental Research
Foundation, Grant No.96-02-19184 and by the INTAS, Grant No.94-2851.}
\\
{\it Institute of
Theoretical and Experimental Physics,}\\
{\it B.Cheremushkinskaya, 25, 117 218, Moscow, Russia}\\
\\
D.EBERT \thanks{E-mail: debert@qft2.physik.hu-berlin.de}\\
{\it Institut f\"ur Elementarteilchenphysik, Humboldt-Universit\"at,}\\
{\it Invalidenstrasse 110, D-10115, Berlin, FRG}\\
\\
YU.A.SIMONOV \thanks{E-mail: simonov@vxitep.itep.ru; supported by the
Russian Fundamental Research Foundation, Grant No.96-02-19184 and by the
INTAS, Grant No.94-2851.} \\
{\it Institute of Theoretical and Experimental Physics,}\\
{\it B.Cheremushkinskaya, 25, 117 218, Moscow, Russia}\\}
\date{}
\begin{document}
\maketitle
\vspace{1cm}

\newcommand{\be}{\begin{equation}}
\newcommand{\ee}{\end{equation}}

\vspace{1cm}
\centerline{\bf {Abstract}}

\vspace{3mm}
Using   cumulant   expansion   for   an   averaged   Wilson   loop  we
derive an action of the gluodynamics string in the form of a series in
powers of the correlation length of the vacuum. In the lowest orders it
contains the Nambu--Goto term and the rigidity term with the  coupling
constants computed from the bilocal correlator of gluonic fields. Some
higher derivative corrections are calculated.

\newpage
{\large \bf 1. Introduction}

\vspace{3mm}
One of the most exciting questions of the modern quantum field theory is a
possible relationship between the large-distance behaviour of the confining
phase  of  gluodynamics  and  string  theory  (for  a  review  see for
example$^{1,2}$). The  aim of  this letter  is to  derive an effective
action for the gluodynamics string using the averaged Wilson loop expressed
in terms  of field  correlators$^{3,4,5,6}$. In  what follows  we keep
for simplicity only the lowest -- bilocal correlators, which are believed to
be dominant according to lattice data$^{5}$, and briefly discuss the effect
of higher correlators. The bilocal correlator may be parametrized in the
following way$^{3,4,5,6}$:

$$<
F_{\mu\nu}(x)\Phi(x,x^{\prime})F_{\lambda\rho}(x^{\prime})\Phi(x^{\prime},x)>
= \frac{\hat{1}}{N_c}\left \{ (\delta_{\mu\lambda}\delta_{\nu\rho}-
\delta_{\mu\rho}\delta_{\nu\lambda})D \Biggl (
\frac{(x-x^{\prime})^2}{T_g^2}\Biggr ) + \right.$$

\be
\left. + \frac{1}{2} \Biggl [ \frac{\partial}{\partial
x_{\mu}}((x-x^{\prime})_{\lambda} \delta_{\nu\rho} -
(x-x^{\prime})_{\rho}\delta_{\nu\lambda}) + \frac{\partial}{\partial x_{\nu}}
((x-x^{\prime})_{\rho} \delta_{\mu\lambda} - (x-x^{\prime})_{\lambda}
\delta_{\mu\rho})\Biggr ] D_1 \Biggl ( \frac{(x-x^{\prime})^2}{T_g^2} \Biggr
) \right \}.
\ee
Here $T_g$ is the correlation length of the vacuum, which is small in the
comparison with the size $r$ of a Wilson loop in the confining
regime$^{7}$.

In order to derive an action for the gluodynamics string, induced by the
nonperturbative background fields, we shall consider the expression for an
averaged Wilson loop using the nonabelian Stokes theorem$^{3}$

\be
<W(C)> = tr~exp \Biggl ( - \frac{g^2}{2} \int\limits_S d\sigma_{\mu\nu}(x)
\int\limits_S
d\sigma_{\lambda\rho}(x^{\prime})<F_{\mu\nu}(x)\Phi(x,x^{\prime})
F_{\lambda\rho}(x^{\prime}) \Phi(x^{\prime},x)> \Biggr ).
\ee
At this point one should make clear of the notion of the surface $S$
entering the integral in (2). In gluodynamics the Wilson loop average
depends on the contour $C$  and should not depend on the shape of the
surface $S$  bounded by this contour, and therefore one usually considers $S$
to be the surface of minimal area. In what follows we generalize this
definition, taking $S$ to be any surface, since our aim is to derive an
effective string action for the case when the surface, swept by this string,
is arbitrary. Such an effective action contains all geometrical
characteristics of the surface in question, and one can always specify the
surface $S$ in the final expression.

To derive this action we shall expand the integral in (2) in powers of
$\frac{T_g}{r}$. Such an expansion yields in the lowest, second, order in
$T_g$ the usual Nambu-Goto term with the string tension proportional to the
surface integral of the function $D$ in agreement with$^{5}$, while
in the order $T_g^4$  there arises the rigidity term$^{8,9}$  with the
inverse bare coupling constant proportional to the first moment of the
function $2D_1-D.$ The sign of this coupling constant is connected with the
type of dual superconductor, describing the nonperturbative gluodynamics
vacuum$^{10}$, and hence we may quote the following result of the next
section: if $\int d^2zz^2(2D_1(z^2) - D(z^2)) < 0$  than this is a type-II
dual superconductor (in the Abelian case the stability of the classical
Abrikosov-Nielsen-Olesen strings is ensured only in the case when this type
of superconductor is realized$^{10}$).

The main results of the letter and possible future developments are
discussed in the Conclusion.

In the Appendix, as an example of geometrical structures arising in higher
orders in $\frac{T_g}{r}$, we present some of the higher derivative terms in
the order $T_g^6$. The corresponding inverse bare coupling constants are
proportional to the second moment of the function $D+D_1$. Therefore the
effective action of the gluodynamics string, generated by the nonperturbative
confining background fields, has the form of a series in powers of
$\frac{T_g}{r}$, corresponding to the series in powers of the scalar
curvature of the manifold and more complicated geometrical structures.

\vspace{6mm}
{\large \bf 2. An action up to the order $T_g^4$}

\vspace{3mm}
Let us rewrite the correlator (1)  in the form, which is more convenient for
the future calculations:

$$< F_{\mu\nu}(x) \Phi(x,x^{\prime}) F_{\lambda\rho}(x^{\prime})
\Phi(x^{\prime},x)> = \frac{\hat{1}}{N_c} \left \{
(\delta_{\mu\lambda}\delta_{\nu\rho} -
\delta_{\mu\rho}\delta_{\nu\lambda}) \Biggl [ D \Biggl (
\frac{(x-x^{\prime})^2}{T_g^2} \Biggr ) + \right. $$

$$+ D_1 \Biggl ( \frac{(x-x^{\prime})^2}{T_g^2} \Biggr ) \Biggr ] +
\frac{1}{T_g^2} [ (x-x^{\prime})_{\mu}(x-x^{\prime})_{\lambda}
\delta_{\nu\rho} - (x-x^{\prime})_{\mu}(x-x^{\prime})_{\rho}
\delta_{\nu\lambda} + (x-x^{\prime})_{\nu} (x-x^{\prime})_{\rho}
\delta_{\mu\lambda} - $$

$$\left. - (x-x^{\prime})_{\nu} (x-x^{\prime})_{\lambda} \delta_{\mu\rho}]
D_1^{\prime}\Biggl ( \frac{(x-x^{\prime})^2}{T_g^2}\Biggr ) \right \},$$
where $D_1^{\prime}$ denotes the derivative of the function $D_1$ by the
argument.

Our first goal is to compute the integral

$$ J \equiv \int d\sigma_{\mu\nu}(x)\int d\sigma_{\lambda\rho}(x^{\prime})
(\delta_{\mu\lambda}\delta_{\nu\rho} - \delta_{\mu\rho}\delta_{\nu\lambda})
\Biggl [ D\Biggl ( \frac{(x-x^{\prime})^2}{T_g^2} \Biggr ) + D_1
\Biggl ( \frac{(x-x^{\prime})^2}{T_g^2} \Biggr )  \Biggr ].$$
Here $d\sigma_{\mu\nu}(x)=\sqrt{g(\xi)}t_{\mu\nu}(\xi)d^2\xi, ~ t_{\mu\nu}=
\frac{1}{\sqrt{g}}\varepsilon^{ab}(\partial_a x_{\mu})(\partial_b x_{\nu})$
is the extrinsic curvature of the string world sheet, $t_{\mu\nu}^2=2,$
$g_{ab}=(\partial_a x_{\mu})(\partial_b x_{\mu})$ is the induced metric
tensor, $g=det\parallel g_{ab} \parallel, \partial_a \equiv $\\
$\equiv \frac{\partial}{\partial \xi^a}; a,b = 1,2$.  Expanding
$\sqrt{g(\xi^{\prime})}, ~t_{\lambda\rho}(\xi^{\prime}),~ x^{\prime}-x$ and
$D\Biggl ( \frac{(x-x^{\prime})^2}{T_g^2} \Biggr ) + D_1 \Biggl (
\frac{(x-x^{\prime})^2}{T_g^2} \Biggr )$, \\
where $x^{\prime} \equiv
x(\xi^{\prime})$, systematically in powers of $\frac{T_g}{r}$, passing from
the ordinary derivatives to the covariant ones (without torsion) via the
familiar Gauss-Weingarten formulae $D_aD_b x_{\mu} = \partial_a\partial_b
x_{\mu}- $\\
$- \Gamma^c_{ab}\partial_c x_{\mu}=K^i_{ab}n_{i\mu},~
n_{i\mu}n_{j\mu}=\delta_{ij},~ n_{i\mu}\partial_a x_{\mu}=0;~ i,j=1,2$, where
$\Gamma^c_{ab}$ is a Christoffel symbol, $K^i_{ab}$ is the second
fundamental form of the manifold, $n_{i\mu}$  are the unit normals to the
sheet, one gets in the conformal gauge $g_{ab}=e^{\varphi}\delta_{ab}$

\be
J = T_g^2 \int d^2\xi\sqrt{g} \Biggl [ 4 M_0 - T_g^2 \frac{M_1}{4}
g^{ab}(\partial_a t_{\mu\nu})(\partial_b t_{\mu\nu}) \Biggr ] + O \Biggl (
\frac{T_g^6 < F^2>}{r^2} \Biggr ),
\ee
where $M_0\equiv\int d^2 z (D(z^2) + D_1(z^2)),$ $M_1\equiv\int d^2 z
z^2(D(z^2)+D(z^2))$, $< F^2 >\equiv$ \\
$\equiv tr~<F_{\mu\nu}(0) F_{\mu\nu}(0)>.$
Here we omitted the full derivative terms of the form $\int
d^2\xi\sqrt{g}R$, where $R$ is a scalar curvature of the manifold, and used
the formula $(D_a D^a x_{\mu})(D_bD^b
x_{\mu})=$ \\
$=g^{ab}(\partial_at_{\mu\nu})(\partial_b t_{\mu\nu}).$  The
estimate  for the neglected terms may be easily obtained if one assumes that
the string world sheet is not much crumpled, so that the induced metric is a
smooth function, which means that the typical values of $\xi$  are of the
order of $r$.

Using the relations $t_{\mu\lambda} t_{\nu\lambda}=g^{ab}(\partial_a
x_{\mu})(\partial_b x_{\nu}),~(g^{ab} g^{cd} + g^{ac} g^{bd} + g^{ad}
g^{bc})(\partial_a t_{\mu\nu})(\partial_b t_{\mu\lambda})\cdot$\\
$\cdot (\partial_c
x_{\lambda})(\partial_d x_{\nu})=K^{b i}_a K^{ai}_b-R$ and omitting the full
derivative terms one can in analogous way compute the integral

$$ \frac{1}{T_g^2} \int d\sigma_{\mu\nu}(x) \int
d\sigma_{\lambda\rho}(x^{\prime}) [(x-x^{\prime})_{\mu}
(x-x^{\prime})_{\lambda} \delta_{\nu\rho}-(x-x^{\prime})_{\mu}
(x-x^{\prime})_{\rho}\delta_{\nu\lambda}+$$

$$+ (x-x^{\prime})_{\nu}(x-x^{\prime})_{\rho}\delta_{\mu\lambda} -
(x-x^{\prime})_{\nu}(x-x^{\prime})_{\lambda} \delta_{\mu\rho} ]
D^{\prime}_1 \Biggl ( \frac{(x-x^{\prime})^2}{T_g^2}\Biggr ),$$
which occurs to be equal to

\be
T_g^2 \int d^2 \xi \sqrt{g} \Biggl [ - 4M_0^{(1)} + T_g^2 \frac{3M^{(1)}_1}
{4} g^{ab}(\partial_a t_{\mu\nu})(\partial_b t_{\mu\nu}) \Biggr ] + O \Biggl
( \frac{T_g^6<F^2>}{r^2}\Biggr ),
\ee
where $M^{(1)}_0\equiv\int d^2 z D_1(z^2)$, $ M^{(1)}_1\equiv\int d^2 z
z^2 D_1(z^2).$

During the derivation of the formulae (3)  and (4) we exploited the fact that
for any odd $n~\int d^2 \xi \xi^{i_1}... \xi^{i_n} D(\xi^2)=\int d^2 \xi
\xi^{i_1}... \xi^{i_n} D_1(\xi^2)=0$.

Combining together (3) and (4) we finally obtain the effective action of
the gluodynamics string, induced by the nonperturbative background fields,
in the approximation when all the correlators higher than bilocal are
neglected:

\be
S_{biloc.} = \frac{g^2}{2} \Biggl [ \sigma \int d^2 \xi \sqrt{g} +
\frac{1}{\alpha_0} \int d^2\xi \sqrt{g}g^{ab}(\partial_a
t_{\mu\nu})(\partial_b t_{\mu\nu}) + O \Biggl ( \frac{T_g^6 < F^2 >}{r^2}
\Biggr ) \Biggr ],
\ee
where

\be
\sigma \equiv 4T_g^2\int d^2z D(z^2)
\ee
is a string tension (which agrees with$^5$) and

\be
\frac{1}{\alpha_0} \equiv \frac{1}{4} T_g^4 \int d^2 z z^2 (2 D_1(z^2) -
D(z^2))
\ee
is an inverse bare coupling constant of the rigidity term.

Hence we proved the statement, announced in the Introduction, namely when
$\int d^2 z z^2(2 D_1(z^2) - D(z^2))<0$, the nonperturbative Euclidean
gluodynamics vacuum may be considered as
a type--II dual superconductor, which in the Abelian Higgs Model case
implies that the Londons$^{\prime}$ penetration depth of magnetic field is
larger than the correlation radius of the Higgs field fluctuations, and the
classical Abrikosov-Nielsen-Olesen strings are stable$^{10}$. In
particular one concludes that when $D>2D_1$ everywhere, the confining regime
of an averaged Wilson loop is realized according to the dual Meissner effect
mechanism$^{11}$ with the string tension given by the formula (6).

In the higher orders in $\frac{T_g}{r}$ more and more complicated geometrical
structures, containing higher covariant derivatives, arise in the string
action. The inverse bare coupling constants of these terms are linear
combinations of higher moments of the functions $D$ and $D_1$. One can
therefore establish some type of correspondence between
the expansion of the string action in powers of $\frac{T_g}{r}$ and a
multipole expansion in two-dimensional gravity. A generic $n$-th $(n\geq 2)$
term of the string action is proportional to some linear combination of the
$(n-1)$-th moments of the functions $D$  and $D_1$, which are of the order
of $T_g^{2n}<F^2>$  and to the $(2n-2)$-th derivative of the induced metric.
Integrating over $d^2\xi$ one obtains an estimate
$<F^2>r^4(\frac{T_g}{r})^{2n}$. As an example we present some geometrical
structures, arising in the string action in the order $T_g^6$, in the
Appendix.

The same effect will be due to the higher correlators. However there is one
more parameter in the problem, which is assumed to be much less than 1 in
the confining regime of the Wilson loop. This is the parameter of cumulant
expansion$^{3}$, $g<F^2>^{\frac{1}{2}}T_g^2$, whose $n$-th power estimates
the upper limit of the $n$-th order cumulant $g^n\ll
F_{\mu_1\nu_1}(x_1)\Phi(x_1,x_2)F_{\mu_2\nu_2}(x_2)...$
$F_{\mu_n\nu_n}(x_n)\Phi(x_n,x_1)\gg$. One may conclude that the higher
terms of cumulant expansion will contribute to $\sigma$ and $\alpha_0$  also
(as well as to the coupling constants of the terms arising in higher orders
of $\frac{T_g}{r}$), but their contributions will be suppressed by the
additional powers of the parameter of cumulant expansion in comparison with
the formulae (6) and (7) derived from the bilocal correlator.

Therefore we see that the Wilson loop average, written through the field
correlators, while expanded in powers of $\frac{T_g}{r}$  gives  rise to the
expansion of the string effective action, which may be called curvature
expansion.

Note that the formulae (3)-(7) (as well as (A.1)-(A.4) presented in the
Appendix) were derived in  the conformal gauge, while we dealt with the open
string, sweeping the area inside the Wilson loop. It is known$^{1}$
that in the case of a unit disc (onto which the string world sheet in our
case may be unambiguously mapped) this gauge is accessible, but one should
take into account diffeomorphisms reparametrizing the boundary. This
reparametrization, defined modulo $SL(2,{\bf R})$  transformations, is
determined by the original metric $g_{ab}(\xi)$, and thus if we quantize the
Nambu-Goto term using the method suggested in$^{12}$, in the
functional integral over all the metrics one should take into account not
only $\varphi$-integration, but also integration over all possible
reparametrizations.

\vspace{6mm}
{\large \bf 3. Conclusion}

\vspace{3mm}
In this letter we used the Wilson loop average, written via the field
correlators, to derive the effective action for the gluodynamics string in
the approximation when all the  correlators higher than bilocal are
neglected. Such an action has the form of a series in powers of the vacuum
correlation length and in the lowest orders is given by the formula (5),
containing the Nambu-Goto term with the string tension, defined via (6), and
the rigidity term with the inverse bare coupling constant (7)  proportional
to the linear combination of the first moments of the functions
parametrizing the bilocal correlator. It is shown that in agreement with the
't Hooft-Mandelstam mechanism of confinement the confining regime of the
Wilson loop takes place in the case of type-II dual superconductor model of
the gluodynamics vacuum, which in the Abelian case corresponds to the
situation when stable Abrikosov-Nielsen-Olesen strings exist. Therefore the
criterion of distinguishing of the confining and deconfining regimes (or the
types of superconductor in the Abelian case), following from the bilocal
correlator, is established. In general expansions of the Wilson loop
average, expressed in terms of field correlators, in powers of the vacuum
correlation length and in powers of the parameter of cumulant expansion
produce the curvature expansion of the effective action of the gluodynamics
string generated by the nonperturbative background fields.

However the approach suggested for derivation of this action leaves us with
the conformal anomaly of the Nambu-Goto term if $D\not=26$. One of the
possible methods of its cancellation in \\
$D=4$  was suggested in$^{13}$
for the case of the strings in the Abelian Higgs Model. In the framework of
this method passing from the field variables to the collective string ones
and computing the Jackobian, corresponding to such transformation, one gets
from it the Polchinski-Strominger term$^{14}$  in the action, which
exactly cancels the conformal anomaly in $D=4$. It should be emphasized that
the Polchinski-Strominger terms do not arise during our derivation of the
effective action (5), and hence one should think about the mechanisms of
cancellation of the conformal anomaly in $D=4$ possibly similar to one
suggested in$^{13}$. While in this letter we derived the effective
action of the string, induced only by the nonperturbative confining fields,
it seems natural to try to disentangle this problem in a rather elegant way
taking into account perturbative gluons$^{\prime}$ contributions and
reformulating the summation over the surfaces in the functional
integral$^{1,12}$ in terms of the perturbative theory in the
nonperturbative gluodynamics background$^{6}$. Within this approach
the perturbative gluons and ghosts, propagating inside the Wilson loop,
generate the string world sheet excitations. The investigation of this
problem will be the topic of another publication.

One more  set of questions is connected with the rigidity term in (5). It is
known$^{1,8}$, that in the generic case the rigid string has a
crumpled world sheet, and its spectrum contains bosonic tachyon -- the
particle with imaginary mass. However as was discussed in$^{1,8}$,
these problems disappear if the $\beta$-function has a zero at some value of
the coupling constant, which may take place for example due to some
$\theta$-terms in the action. This problem will be also  treated elsewhere.

\vspace{6mm}
{\large \bf 4. Acknowledgements}

\vspace{3mm}
D.A. and Yu.S. are grateful to A.R.Kavalov for useful discussions and
M.Markina for typing the manuscript. D.A. would like to thank Professor
D.Ebert and all the staff of the Quantum Field Theory Department of the
Institut f\"ur Physik of the Humboldt-Universit\"at  of Berlin, where the
initial part of this work was done, for kind hospitality.

\vspace{6mm}
{\large \bf Appendix.~~ Some examples of geometrical structures arising in
the action in the order $T_g^6$.}

\setcounter{equation}{0}
\def\theequation{A.\arabic{equation}}

\vspace{3mm}
In this Appendix we present a part of the string
action terms in the next order in $T_g$. Namely we demonstrate the
geometrical structures arising from $J$ in the order $T_g^6$.

One can prove that $O\Biggl ( \frac{T_g^6<F^2>}{r^2} \Biggr )$  in (3)
equals to

$$T_g^6 \int d^2 \xi \sqrt{g}d^2\lambda \left \{ J_1+J_2+J_3+J_4+P(\lambda^2)
\Lambda^{abcd}\Biggl [ \frac{1}{6}(\partial_a\partial_b\partial_c\partial_d
\sqrt{g}) - \frac{1}{2}(\partial_a t_{\mu\nu})(\partial_b
t_{\mu\nu})(\partial_c\partial_d\sqrt{g}) + \right.$$
$$+ \frac{1}{3}t_{\mu\nu}(\partial_a\partial_b\partial_ct_{\mu\nu})
(\partial_d\sqrt{g}) + \frac{1}{12}\sqrt{g}t_{\mu\nu}
(\partial_a\partial_b\partial_c\partial_d t_{\mu\nu}) \Biggr ] +
P^{\prime}(\lambda^2)\Lambda^{abcdef}\Biggl [ \frac{1}{3}(\partial_a\sqrt{g})
(2(\partial_b\partial_c x_{\mu}) \cdot$$
$$\cdot (\partial_d\partial_e\partial_f x_{\mu}) + (\partial_b x_{\mu})
(\partial_c\partial_d\partial_e\partial_f x_{\mu})) +
(\partial_a\partial_b\sqrt{g}) \Biggl ( \frac{1}{2}(\partial_c\partial_d x_{\mu})
(\partial_e\partial_f x_{\mu}) + \frac{2}{3}(\partial_c x_{\mu})
(\partial_d\partial_e\partial_f x_{\mu})\Biggr ) + $$
$$+ \frac{2}{3}(\partial_a\partial_b\partial_c\sqrt{g})(\partial_dx_{\mu})
(\partial_e\partial_f x_{\mu}) -
\sqrt{g}(\partial_at_{\mu\nu})(\partial_bt_{\mu\nu})\Biggl(\frac{1}{4}
(\partial_c\partial_dx_{\lambda})(\partial_e\partial_fx_{\lambda})+\frac{1}{3}
(\partial_cx_{\lambda})(\partial_d\partial_e\partial_fx_{\lambda})
\Biggr ) - $$
$$-(\partial_at_{\mu\nu})(\partial_bt_{\mu\nu})(\partial_c\sqrt{g})
(\partial_dx_{\lambda})(\partial_e\partial_fx_{\lambda})
+ \frac{1}{3} \sqrt{g}t_{\mu\nu}(\partial_a\partial_b\partial_c
t_{\mu\nu}) (\partial_dx_{\lambda})(\partial_e\partial_fx_{\lambda})\Biggr]+
P^{\prime\prime}(\lambda^2)\Lambda^{abcdefij}\cdot$$
$$\cdot \Biggl
[(\partial_a\sqrt{g})(\partial_bx_{\mu})(\partial_c\partial_dx_{\mu})
\Biggl ((\partial_e\partial_f x_{\nu})(\partial_i\partial_j
x_{\nu})+\frac{4}{3} (\partial_e
x_{\nu})(\partial_f\partial_i\partial_jx_{\nu})\Biggr ) +
(\partial_a\partial_b\sqrt{g})(\partial_cx_{\mu})
(\partial_d\partial_ex_{\mu})(\partial_fx_{\nu})\cdot$$
$$\cdot (\partial_i\partial_jx_{\nu})-\frac{1}{2}\sqrt{g}(\partial_a
t_{\mu\nu}) (\partial_b t_{\mu\nu})(\partial_cx_{\lambda})(\partial_d
\partial_ex_{\lambda})(\partial_fx_{\rho})(\partial_i\partial_jx_{\rho})
\Biggr]+
\frac{2}{3}P^{\prime\prime\prime}(\lambda^2)\Lambda^{abcdefijkl}
(\partial_a\sqrt{g})\cdot$$
$$\cdot(\partial_bx_{\mu})(\partial_c\partial_dx_{\mu})(\partial_ex_{\nu})
(\partial_f\partial_ix_{\nu})(\partial_jx_{\lambda})
(\partial_k \partial_lx_{\lambda})+\frac{1}{6}\sqrt{g}P^{IV}(\lambda^2)
\Lambda^{abcdefijklmn}(\partial_ax_{\mu})(\partial_b\partial_cx_{\mu})
(\partial_dx_{\nu})\cdot $$
$$\cdot (\partial_e\partial_fx_{\nu})(\partial_ix_{\lambda})
(\partial_j \partial_kx_{\lambda})(\partial_lx_{\rho})
(\partial_m\partial_nx_{\rho}) \Biggr \}.$$
Here $P \equiv D+D_1, \lambda^2\equiv g_{ab}\lambda^a\lambda^b,
\Lambda^{i_1...i_n} \equiv \lambda^{i_1}... \lambda^{i_n}$, and
$J_1,J_2,J_3,J_4$  are the following terms, which do not contain derivatives
of $P$ higher than of the third order and do not depend explicitly  on
$\sqrt{g}t_{\mu\nu}, \sqrt{g}\partial_at_{\mu\nu}$ and derivatives of
$\sqrt{g}$:
$$J_1\equiv\frac{1}{3}\sqrt{g}P^{\prime}(\lambda^2)\Lambda^{abcdef}
\Biggl (\frac{1}{2}(\partial_a\partial_bx_{\mu})
(\partial_c\partial_d\partial_e\partial_fx_{\mu})
+\frac{1}{3}
(\partial_a\partial_b\partial_cx_{\mu})(\partial_d\partial_e\partial_fx_{\mu})
+\frac{1}{5}(\partial_ax_{\mu})
(\partial_b\partial_c\partial_d\partial_e\partial_fx_{\mu})\Biggr),$$
$$J_2\equiv 2\sqrt{g}P^{\prime\prime}(\lambda^2)\Lambda^{abcdefij}
\Biggr(\frac{1}{4}(\partial_a\partial_bx_{\mu})
(\partial_c\partial_dx_{\mu})
+\frac{1}{3}
(\partial_ax_{\mu})(\partial_b\partial_c\partial_dx_{\mu})\Biggr)\cdot$$
$$\cdot\Biggl(\frac{1}{4}
(\partial_e\partial_fx_{\nu})
(\partial_i\partial_jx_{\nu})+\frac{1}{3}(\partial_ex_{\nu})
(\partial_f\partial_i\partial_jx_{\nu})\Biggr),$$
$$J_3\equiv \frac{1}{3}\sqrt{g}P^{\prime\prime}(\lambda^2)\Lambda^{abcdefij}
(\partial_ax_{\mu})(\partial_b\partial_cx_{\mu})
(2(\partial_d\partial_ex_{\nu})
(\partial_f\partial_i\partial_jx_{\nu})
+(\partial_dx_{\nu})
(\partial_e\partial_f\partial_i\partial_jx_{\nu})),$$
$$J_4\equiv 2\sqrt{g}P^{\prime\prime\prime}(\lambda^2)\Lambda^{abcdefijkl}
(\partial_ax_{\mu})(\partial_b\partial_cx_{\mu})
(\partial_dx_{\nu})
(\partial_e\partial_fx_{\nu})\cdot$$
$$\cdot \Biggl(\frac{1}{4}
(\partial_i\partial_jx_{\lambda})
(\partial_k\partial_lx_{\lambda})+\frac{1}{3}(\partial_ix_{\lambda})
(\partial_j\partial_k\partial_lx_{\lambda})\Biggr).$$
Using the standard rules of computation of higher covariant derivatives and
the Gauss-Weingarten formulae it is possible to calculate the integrals
$\int d^2\lambda J_1,..., \int d^2\lambda J_4.$  The results have the form:
$$\int d^2 \lambda J_1=-\frac{M_2}{48}\left
\{\frac{1}{3}\Biggl [ (D_aD^2x_{\mu})
(2D^2D^a+D^aD^2+2D_bD^aD^b)x_{\mu}+(D^2D_ax_{\mu})(D^2D^a+
2D_bD^aD^b)x_{\mu}+\right.$$
$$+(D_aD_bD^ax_{\mu})D_cD^bD^cx_{\mu}+(D_aD_bD_cx_{\mu})(D^aD^bD^c+D^aD^cD^b+
D^bD^cD^a+D^cD^bD^a+D^cD^aD^b+$$
$$+D^bD^aD^c)x_{\mu}\Biggr]+\frac{1}{2}(D_aD_bx_{\mu})(D^2D^aD^b+D^2D^bD^a+D^aD^2D^b
+D^bD^2D^a+D^aD^bD^2+D^bD^aD^2+$$
$$+D^aD_cD^bD^c
+D^bD_cD^aD^c+D_cD^aD^bD^c+D_cD^bD^aD^c+D_cD^aD^cD^b+D_cD^bD^cD^a)x_{\mu}+$$
$$+\frac{1}{2}(D^2x_{\mu})(D_aD^2D^a+D_aD_bD^aD^b
+D^2D^2)
x_{\mu}+\frac{1}{5}(\partial_ax_{\mu})(D^aD^2D^2+D^2D^aD^2+D^2D^2D^a+$$
$$+D^2D_bD^aD^b+D_bD^2D^aD^b+D^aD_bD^2D^b+D_bD^aD^bD^2
+D_bD^2D^bD^a+D_bD^aD^2D^b+$$
$$+D^aD_bD_cD^bD^c+D_bD^aD_cD^bD^c+D_bD_cD^aD^bD^c+
D_bD_cD^bD^aD^c+D_bD_cD^bD^cD^a
+D_bD_cD^aD^cD^b)x_{\mu}+$$
$$+(\partial_a\varphi)\Biggr[(D^2x_{\mu})\Biggl(\frac{1}{2}D^2D^a
+2D_bD^aD^b+\frac{7}{2}D^aD^2\Biggr)x_{\mu}+(D_bD^2x_{\mu}+D^2D_bx_{\mu}+
D_cD_bD^cx_{\mu})\cdot$$
$$\cdot
(3D^aD^b+D^bD^a)x_{\mu}+(D_bD_cx_{\mu})\Biggl(2D^bD^aD^c+2D^cD^aD^b+
\frac{1}{2}D^bD^cD^a+\frac{1}{2}D^cD^bD^a+\frac{7}{2}D^aD^bD^c+$$
$$+\frac{7}{2}D^aD^cD^b\Biggr)x_{\mu}+\frac{1}{10}(\partial_bx_{\mu})(7D^2D^aD^b
+7D_cD^bD^aD^c+7D^bD_cD^aD^c+19D^aD^2D^b+19D^aD^bD^2+$$
$$+19D^aD_cD^bD^c+13D^bD^aD^2+13D_cD^aD^cD^b+13D_cD^aD^bD^c+D_cD^bD^cD^a
+D^2D^bD^a+D^bD^2D^a)x_{\mu}+$$
$$+(\partial^ax_{\mu})(D_cD_bD^cD^b+D_bD^2D^b+D^2D^2)x_{\mu}\Biggr]-\frac{1}{30}
(D_aD_bx_{\mu})(D_cD^bx_{\mu})(7(\partial^a\varphi)\partial^c\varphi
+13\partial^a\partial^c\varphi)-$$
$$-\frac{1}{30}(D_aD_bx_{\mu})(D^aD_cx_{\mu})((\partial^b\varphi)
\partial^c\varphi+7\partial^b\partial^c\varphi)-(D_aD_bx_{\mu})
(D_cD^ax_{\mu})\Biggl(\frac{9}{5}(\partial^b\varphi)\partial^c\varphi+
\frac{2}{3}\partial^b\partial^c\varphi\Biggr)-$$
$$-\frac{1}{3}(D_aD_bx_{\mu})(D^2x_{\mu})\Biggl(\frac{19}{5}
(\partial^a\varphi)
\partial^b\varphi+2\partial^a\partial^b\varphi\Biggr) -\frac{1}{3}
(D_aD_bx_{\mu})(D^bD^ax_{\mu})\Biggl(2\partial^2\varphi+\frac{89}{10}
(\partial_c\varphi)^2\Biggr)-$$
$$-\frac{1}{3}(D_aD_bx_{\mu})(D^aD^bx_{\mu})\Biggl(2\partial^2\varphi +
\frac{53}{10}(\partial_c\varphi)^2\Biggr)-\frac{1}{3}(D^2x_{\mu})(D^2x_{\mu})
(2\partial^2\varphi+5(\partial_c\varphi)^2)+$$
$$+\sqrt{g}\Biggl(\frac{26}{15}
(\partial_a\partial_b\varphi)^2
+\frac{89}{20}(\partial_a\partial_b\varphi)(\partial^a\varphi)
\partial^b\varphi+\frac{23}{15}(\partial^2\varphi)^2+$$
\be
\left.+\frac{13}{8}
(\partial^2\varphi)(\partial_a\varphi)^2+5(\partial^2\partial_a\varphi)
\partial^a\varphi+\frac{9}{5}\partial^4\varphi+\frac{5}{8}
(\partial_a\varphi)^2(\partial_b\varphi)^2\Biggr)\right\},
\ee
$$\int d^2\lambda J_2=\frac{M_2}{16}\left \{ \frac{1}{72}\Biggl [ (D_aD_b
x_{\mu})(D^2x_{\mu})((D^2
x_{\nu})(D^aD^b+D^bD^a)x_{\nu}+2(D^aD_cx_{\nu}+D_cD^ax_{\nu})(D^bD^c+\right.$$
$$+D^cD^b)
x_{\nu})+
(D_aD_bx_{\mu})(D_cD_dx_{\mu})(2(D^cD^bx_{\nu})D^dD^ax_{\nu}
+2(D^cD^ax_{\nu})D^dD^b
x_{\nu}+(D^aD^bx_{\nu}+D^bD^ax_{\nu})\cdot$$
$$\cdot(D^cD^d+D^dD^c)x_{\nu}+
(D^cD^ax_{\nu})D^bD^dx_{\nu}+(D^aD^cx_{\nu})D^dD^bx_{\nu}+(D^bD^cx_{\nu})
D^dD^ax_{\nu}+$$
$$+(D^cD^bx_{\nu})D^aD^dx_{\nu})+
(D_aD_bx_{\mu}+D_bD_ax_{\mu})(D^bD_cx_{\mu}+D_cD^bx_{\mu})(D^2x_{\nu})
D^cD^ax_{\nu}+$$
$$+(D_aD_bx_{\mu})(D_cD_dx_{\nu}+
D_dD_cx_{\nu})((D^bD^cx_{\mu}
+D^cD^bx_{\mu})(D^aD^d+D^dD^a)x_{\nu}+$$
$$+(D^bD^cx_{\nu}+D^cD^bx_{\nu})(D^aD^d+
D^dD^a)x_{\mu})+\frac{1}{2}(D^2x_{\mu})(D^2x_{\mu})(D_aD_bx_{\nu})\cdot$$
$$\cdot(D^aD^b+D^bD^a)x_{\nu}+(D_aD_bx_{\mu})(D^aD^bx_{\mu})(D_cD_dx_{\nu})
\Biggl(\frac{1}{2}D^cD^d+D^dD^c\Biggr)x_{\nu}+\frac{1}{2}(D_aD_bx_{\mu})(D^bD^ax_{\mu})
\cdot$$
$$\cdot (D_cD_dx_{\nu})(D^dD^cx_{\nu})\Biggr ] -
\frac{1}{9}\sqrt{g}(D^2x_{\mu})(D^2x_{\mu})\Biggl(2\partial^2\varphi+\frac{17}{2}
(\partial_c\varphi)^2\Biggr)-\sqrt{g}(D_aD_bx_{\mu})(D^aD^bx_{\mu}+$$
$$+D^bD^ax_{\mu})
\Biggl ( \frac{2}{9}\partial^2\varphi+\frac{5}{6}
(\partial_c\varphi)^2\Biggr ) - \frac{1}{9}\sqrt{g}\Biggl [ 8(D^2x_{\mu})
(D_aD_bx_{\mu})(\partial^a\partial^b\varphi+(\partial^a\varphi)
\partial^b\varphi)+(\partial_a\partial_b\varphi)\cdot $$
$$\cdot (7(D_cD^ax_{\mu})D^cD^bx_{\mu}+8(D^aD_cx_{\mu})D^cD^bx_{\mu}+
4(D^aD_cx_{\mu})D^bD^cx_{\mu})+(\partial_a\varphi)(\partial_b\varphi)
((D_cD^bx_{\mu})D^cD^ax_{\mu}+$$
$$+6(D_cD^ax_{\mu})D^bD^cx_{\mu}+2(D^aD_cx_{\mu})D^bD^cx_{\mu})\Biggr]+
\frac{1}{3}g\Biggl[8(\partial_a\partial_b\varphi)^2+16
(\partial_a\partial_b\varphi)(\partial^a\varphi)\partial^b\varphi+$$
\be
\left.+4(\partial^2\varphi)(\partial_a\varphi)^2+4(\partial^2\varphi)^2+
\frac{91}{8}(\partial_a\varphi)^2(\partial_b\varphi)^2 \Biggr]\right\},
\ee
$$\int d^2\lambda J_3=\frac{M_2}{48}\sqrt{g}\Biggl \{ (\partial_a\varphi)
[4(D_bD_cx_{\mu})(D^aD^bD^c+D^bD^aD^c+D^cD^aD^b+D^aD^cD^b+D^bD^cD^a+$$
$$+D^cD^bD^a)x_{\mu}+(D^aD_bx_{\mu}+D_bD^ax_{\mu})(4D^2D^b+4D_cD^bD^c+D^bD^2)
x_{\mu}+(D^2x_{\mu})(4D^2D^a+4D_cD^aD^c+$$
$$+D^aD^2)x_{\mu}]+(\partial_ax_{\mu})\Biggl[2(\partial_b\varphi)\Biggl(D_cD^aD^bD^c
+D_cD^bD^aD^c+D_cD^bD^cD^a+D^bD_cD^aD^c+$$
$$+D^aD_cD^bD^c+D_cD^aD^cD^b+D^2D^aD^b+D^2D^bD^a+\frac{5}{4}D^aD^bD^2+
\frac{5}{4}D^bD^aD^2\Biggr)x_{\mu}+$$
$$+(\partial^a\varphi)\Biggl(2D_bD^2D^b+2D_bD_cD^bD^c
+\frac{1}{2}D^2D^2\Biggr)x_{\mu}\Biggr]+
\sqrt{g}[17(\partial^2\varphi)(\partial_a\varphi)^2+7(\partial_a\varphi)^2
(\partial_b\varphi)^2+$$
$$+9(\partial_a\varphi)\partial^2\partial^a\varphi+12(\partial_a\partial_b
\varphi)(\partial^a\varphi)\partial^b\varphi]-2(\partial_a\varphi)
(\partial_b\varphi)[(D^bD_cx_{\mu}+D_cD^bx_{\mu})(D^aD^c+D^cD^a)x_{\mu}+$$
\be
+(D^2x_{\mu})(D^aD^b+D^bD^a)x_{\mu}]-2(\partial_c\varphi)^2[(D_aD_bx_{\mu}
+D_bD_ax_{\mu})(D^aD^b+D^bD^a)x_{\mu}+2(D^2x_{\mu})D^2x_{\mu}]
\Biggr\},
\ee
$$\int d^2\lambda J_4 = \frac{M_2}{32}g \left \{
\frac{1}{3}(\partial_a\varphi)(\partial_b\varphi)[13(D^aD^bx_{\mu})D^2x_{\mu}
+11(D^aD_cx_{\mu})D^cD^bx_{\mu}+7(D^aD_cx_{\mu})D^bD^cx_{\mu}+\right.$$
$$+7(D_cD^ax_{\mu})D^cD^bx_{\mu}]+\frac{1}{3}(\partial_c\varphi)^2 \Biggl [
\frac{9}{2}(D^2x_{\mu})D^2x_{\mu}+4(D_aD_bx_{\mu})D^aD^bx_{\mu}+2(D_aD_b
x_{\mu})D^bD^ax_{\mu}\Biggr]-$$
\be
-2\sqrt{g}\Biggl[12(\partial_a\partial_b\varphi)(\partial^a\varphi)
\partial^b\varphi +
\frac{355}{24}(\partial_a\varphi)^2(\partial_b\varphi)^2+
\frac{17}{3}(\partial^2\varphi)(\partial_a\varphi)^2\Biggr] \Biggr\},
\ee
where $M_2\equiv \int
d^2z(z^2)^2(D(z^2)+D_1(z^2)),~D^2\equiv D_iD^i,~
\partial^2\equiv\partial_i\partial^i,~$
$\partial^4\equiv\partial_i\partial^i\partial_j\partial^j,$\\
$(\partial_a\varphi)^2\equiv(\partial_a\varphi)(\partial^a\varphi),~
(\partial_a\partial_b\varphi)^2\equiv
\partial_a\partial_b\partial^a\partial^b\varphi.$

\newpage
{\large\bf References}

\vspace{3mm}
\noindent
1.~ A.M.Polyakov, {\it Gauge Fields and Strings} (Harwood, 1987).\\
2.~ J.Polchinski, {\it hep-th} /9210045.\\
3.~ Yu.A.Simonov, {\it Yad.Fiz.} {\bf 50}, 213 (1989).\\
4.~ H.G.Dosch, {\it Phys.Lett.} {\bf B190}, 177 (1987); Yu.A.Simonov, {\it
Nucl.Phys.} {\bf B307},

512 (1988); H.G.Dosch and Yu.A.Simonov, {\it Phys.Lett.} {\bf B205}, 339
(1988),

{\it Z.Phys.} {\bf C45}, 147 (1989); Yu.A.Simonov, {\it Nucl.Phys.} {\bf
B324}, 67 (1989),

{\it Phys.Lett.} {\bf B226}, 151 (1989), {\it Phys.Lett.} {\bf B228}, 413
(1989).\\
5.~ Yu.A.Simonov, {\it Yad.Fiz.} {\bf 54}, 192 (1991).\\
6.~ Yu.A.Simonov, {\it Yad.Fiz.} {\bf 58}, 113 (1995).\\
7.~ M.Campostrini et al., {\it Z.Phys.} {\bf C25}, 173 (1984); A. Di Giacomo
and

H.Panagopoulos, {\it Phys.Lett.} {\bf B285}, 133 (1992); I.J.Ford et al.,
{\it Phys.Lett.}

{\bf B208}, 286 (1988); E.Laermann et al., {\it Nucl.Phys.} {\bf B26}
(Proc. Suppl.), 268 (1992).\\
8.~ A.M.Polyakov, {\it Nucl.Phys.} {\bf B268}, 406 (1986).\\
9.~ H.Kleinert, {\it Phys.Lett.} {\bf B174}, 335 (1986); T.L.Curtright et
al., {\it Phys.Rev.}

{\bf D34}, 3811 (1986); G.Germ\'an. {\it Mod.Phys.Lett.} {\bf A6}, 1815
(1991).\\
10. P.Orland, {\it Nucl.Phys.} {\bf B428}, 221 (1994).\\
11. G.'t Hooft, in {\it High Energy Physics}, ed. A.Zichichi (Editrice
Compositori, 1976);

S.Mandelstam, {\it Phys.Lett.} {\bf B53}, 476  (1975).\\
12. A.M.Polyakov, {\it Phys.Lett.} {\bf B103}, 207 (1981).\\
13. E.T.Akhmedov et al., {\it Phys.Rev.} {\bf D53}, 2087 (1996).\\
14. J.Polchinski and A.Strominger, {\it Phys.Rev.Lett.} {\bf 67}, 1681
(1991).
\end{document}